\documentclass[twocolumn,floatfix,superscriptaddress,a4paper,showpacs,showkeys,reprint,prc,nofootinbib]{revtex4-2}
\usepackage{epsfig}
\usepackage{longtable}
\usepackage{latexsym}
\usepackage[colorlinks=true,linktocpage=true,linkcolor=blue,citecolor=blue,allcolors=blue]{hyperref}
\usepackage{url}
\usepackage[utf8]{inputenc}
\usepackage{enumerate}
\usepackage{color}
\usepackage{xcolor}
\usepackage{setspace}
\usepackage{amsmath}
\usepackage{amssymb}
\usepackage[english]{babel}
\usepackage{url}
\usepackage{footnote}
\usepackage{soul}
\usepackage{placeins}
\begin{document}

 \title{Toward a foundation model for heavy-ion collision experiments \\ based on point-cloud diffusion}

\author{Manjunath Omana Kuttan}
\email{manjunath@fias.uni-frankfurt.de}
\affiliation{Frankfurt Institute for Advanced Studies, Ruth-Moufang-Str. 1, D-60438 Frankfurt am Main, Germany}

\affiliation{Xidian-FIAS International Joint Research Center, Giersch Science Center, D-60438 Frankfurt am Main, Germany}

\author{Kai Zhou}
\email{zhoukai@cuhk.edu.cn}
\affiliation{School of Science and Engineering, The Chinese University of Hong Kong, Shenzhen, P.R. China}

\affiliation{Frankfurt Institute for Advanced Studies, Ruth-Moufang-Str. 1, D-60438 Frankfurt am Main, Germany}

\author{Jan Steinheimer}
\affiliation{GSI Helmholtzzentrum f\"ur Schwerionenforschung GmbH, Planckstr. 1, D-64291 Darmstadt, Germany}

\affiliation{Frankfurt Institute for Advanced Studies, Ruth-Moufang-Str. 1, D-60438 Frankfurt am Main, Germany}

\author{Horst Stoecker}
\affiliation{Frankfurt Institute for Advanced Studies, Ruth-Moufang-Str. 1, D-60438 Frankfurt am Main, Germany}

\affiliation{GSI Helmholtzzentrum f\"ur Schwerionenforschung GmbH, Planckstr. 1, D-64291 Darmstadt, Germany}
\affiliation{Institut f\"{u}r Theoretische Physik, Goethe Universit\"{a}t Frankfurt, Max-von-Laue-Str. 1, D-60438 Frankfurt am Main, Germany}
\date{\today}
\begin{abstract}
A novel point cloud diffusion model for relativistic heavy-ion collisions, capable of ultra-fast generation of complete, event-by-event collision output, is introduced.  When trained on UrQMD cascade simulations, the model generates realistic collision event output containing 26 distinct hadron species, as a list of  particle momentum vectors along with their particle ID. From solving inverse problems to accelerating model calculations or detector simulations, the model can be a promising general purpose tool for heavy-ion collisions beneficial to both theoretical studies and experimental applications.

\end{abstract}

\maketitle
Relativistic nucleus-nucleus collisions create short-lived, strongly interacting systems characterized by high temperatures and/or densities, providing a unique terrestrial experimental probe of the conjectured phase diagram of quantum chromodynamics (QCD)\cite{Stoecker:1986ci,Fukushima:2010bq,Rajagopal:2000wf,Halasz:1998qr}. Although several observables \cite{Stoecker:1986ci,Hofmann:1976dy,Stoecker:2004qu,Stephanov:1998dy,Hatta:2003wn,Zhou:2014kka} have been shown to be sensitive to the properties and dynamics of the matter created in these collisions, essential details of the phase diagram, including the exact locations of phase boundaries, the nature of the transitions between them, and the presence of any possible critical points, remain poorly understood. Currently, first principle lattice QCD calculations \cite{Aoki:2006we, Borsanyi:2010cj, HotQCD:2014kol,Cheng:2007jq,Bazavov:2011nk,PACS-CS:2008bkb} are possible only at vanishing and small net-baryon densities, necessitating QCD inspired effective models alongside direct experimental searches to accurately map the high net-baryon density regions of the QCD phase diagram. 

Numerous experimental programs are currently dedicated to investigate QCD matter at high baryon density \cite{HADES:2009aat,Yang:2017llt, STAR:2010vob, STAR:2017sal,Meehan:2016qon,CBM:2016kpk,Yang:2013yeb,Golovatyuk:2016zps}. These experiments will generate copious amounts of high precision data at beam energies of $\sqrt{s_{\mathrm{NN}}} \leq $15~GeV. This wealth of data offers a unique opportunity to investigate baryon rich QCD matter in previously uncharted ways, employing rare and novel probes. 

Significant progress has also been made in modeling moderate to intermediate energy collisions with various approaches emerging. Transport models \cite{Bass:1998ca,Bleicher:1999xi,Cassing:2009vt,Nara:1999dz,Buss:2011mx,Lin:2004en} provide a microscopic, non-equilibrium description of the collisions, while hybrid models \cite{Petersen:2008dd} incorporate a hydrodynamic description for the hot, dense, intermediate phase that depends on the unknown high density and temperature equation of state (EoS). Extensive efforts have also been made to implement realistic density and momentum  dependent EoSs within transport descriptions \cite{OmanaKuttan:2022the,Steinheimer:2023xzs,Steinheimer:2024eha,Nara:2020ztb,Oliinychenko:2022uvy}.  

Despite the substantial advancements in realistic descriptions of heavy-ion collisions, these models suffer from their high computational expense. Even the fastest simulation models run orders of magnitude slower than the rate of experimental data collection. This computational bottleneck limits our ability to perform modern model-data analysis and fully capitalize on the data to be produced by these next-generation experiments.

Extracting fundamental properties of the system such as the EoS or the location of a possible critical point in the phase diagram from the experimental data would require comparisons to model calculations for multiple observables such as collective flow, fluctuations and correlations. For multi-parameter fits or Bayesian analysis, fast emulators are developed using machine learning techniques to quickly map  model parameters to relevant observables \cite{Bernhard:2016tnd,JETSCAPE:2020mzn,Pratt:2015zsa,Nijs:2020roc,OmanaKuttan:2022aml,Oliinychenko:2022uvy,Huth:2021bsp}. The main shortcoming of these approaches is that a new emulator needs to be trained for every new observable included in the inference. This would quickly become unfeasible when a large number of observables such as multi differential flow and n-body correlation spectra are to be included in the analysis.  

In addition to model simulations, experiments also rely heavily on expensive detector simulations for physics analyses, detector calibration, efficiency correction, uncertainty estimation, etc. Frameworks like GEANT4 \cite{GEANT4:2002zbu} which are often used to model various experimental effects, can be substantially slower than experimental data rates, limiting their applicability in real-time analyses, online event selection and event characterization. As a result, the computational demands of detector simulations present an additional bottleneck, which poses a significant challenge to fully exploiting experimental data.

Deep Learning (DL) methods have emerged as a novel approach for fast and accurate analysis of experimental data in high energy heavy-ion collisions (HIC) \cite{Zhou:2023pti,OmanaKuttan:2020brq, OmanaKuttan:2020btb, OmanaKuttan:2023bnb, Pang:2016vdc, Du:2019civ,Du:2020pmp,He:2023zin,Li:2020qqn,Mallick:2022alr,Song:2021rmm,Hirvonen:2023lqy,Sergeev:2020fir}. However, these models are often designed only for a specific task (e.g., predicting the flow coefficients) within a given experiment, limiting their flexibility as a versatile AI tool. A more general approach can be considered using deep generative models. As a class of machine learning algorithms capable of learning the underlying distribution of the training data, generative models can produce new data samples, offering a flexible alternative to DL models that perform only a specific task.

A recent work \cite{Sun:2024lgo} concurrent to ours, applied generative models to generate $p_T$ - $\phi$ spectra (as 64 $\times$ 64 pixel image) of charged particles produced in collisions. Similarly, in \cite{Torbunov:2024iki}, it was shown that a simulation chain with an event generator and detector simulation of the sPHENIX \cite{Aidala:2012nz} experiment can be replaced using a generative model that generates the final calorimeter response. In \cite{Huang:2018fzn}, a deep learning model was trained to predict the final energy density and flow velocity profiles in relativistic hydrodynamic simulations. Although these methods are promising, they are currently limited to producing partial information in the form of histograms of the event, lacking the capability to generate complete event-level data, i.e., a complete list of all hadrons emitted from a given event, including information on their 4-momenta and hadron type. Generative models that generate complete event-level data will be able to emulate an entire simulation chain for an experiment with remarkable speed, offering pathways to build a robust and adaptable AI analysis framework for both heavy-ion collision experiments and theory.

In this work, we introduce, for the first time, an event-by-event generative model for heavy-ion collisions, which generates collision events as lists of particle vectors, named \emph{\textbf{HEIDi}: \textbf{H}eavy-ion \textbf{E}vents through \textbf{I}ntelligent \textbf{Di}ffusion}. With its strong ability to generate complete event information in a flexible representation, this model lays the groundwork for a future foundation model for HIC. Foundation models are general-purpose DL models that can be easily adapted for different tasks without requiring extensive retraining. Furthermore, the model’s adaptability extends beyond HIC, offering potential applications in accelerating cascade simulations of cosmic ray showers or detector simulations in particle and astroparticle physics.

In \emph{HEIDi}, we employ a conditional diffusion probabilistic model based on \cite{Luo_2021_CVPR}, to generate a point cloud of final state particles in a collision event. Diffusion probabilistic models \cite{croitoru2023diffusion, yang2023diffusion} are generative models that are inspired by non-equilibrium thermodynamics. It comprises a forward diffusion process and a reverse anti-diffusion process. During the forward diffusion process, the point cloud of final state particles evolves stochastically over several time steps until it becomes indistinguishable from pure noise, akin to a non-equilibrium system in contact with a heat bath. The generation process is then achieved by learning to reverse this diffusion process. Starting from random samples from a simple well-defined distribution, the model iteratively predicts the noise added in the previous step and then denoise it, ultimately transforming it into a clean, structured point cloud. 

Diffusion-based generative models have been applied in high energy physics for various tasks  \cite{Wang:2023exq,Mikuni:2022xry, Amram:2023onf,Acosta:2023zik,Mikuni:2023tqg,Buhmann:2023bwk,Butter:2023fov,Leigh:2023zle,Buhmann:2023zgc,Buhmann:2023pmh,Mikuni:2023dvk,Leigh:2023toe,Araz:2024bom,Devlin:2023jzp}. In particle physics, such point cloud generative models have primarily been used for simulating jets or calorimeter responses and typically involve small point clouds with limited number of particle types. These models are mainly intended for tasks such as performance benchmarking, detector calibration, or algorithm development. Our work addresses a more complex problem of simulating complete heavy-ion events containing over a thousand particles and dozens of hadron species with the additional but important goal of sophisticated downstream tasks such as addressing inverse problems and online event analyses. Given that heavy-ion physics strongly rely on model-to-data comparisons, an easily adaptable and fast, full event generation framework will significantly enhance the physics potential of current and future experiments.

The present work realizes a point cloud diffusion model for heavy-ion collisions by implementing a deep generative emulator of the microscopic UrQMD cascade model \cite{Bass:1998ca, Bleicher:1999xi}. UrQMD is widely used in the description of nucleus-nucleus, hadron-nucleus and hadron-hadron collisions over a wide range of energies. It is based on the covariant propagation of hadrons and provides an effective solution of the relativistic Boltzmann equation, with n-body correlations. 

In this work, as proof of concept, the model was trained to generate Au - Au collision events at $10A$ GeV using 18000 UrQMD events, each with impact parameter $b$=1 fm. Each event output generated by UrQMD is represented as a point cloud $\mathbf{X}^{(0)} = \{ \mathbf{x}^{(0)}_i \}_{i=1}^{1084}$ where each point is a particle described by its momentum vector and particle information (ID) i.e., $ \mathbf{x}_i^{(0)} = \{ \mathbf{p}_i^{(0)}, \text{ID}_i^{(0)} \}, \quad \text{where} \quad \mathbf{p}_i^{(0)} = (p_{x_i}^{(0)}, p_{y_i}^{(0)}, p_{z_i}^{(0)})$. The number of points in an event is fixed to be 1084, a number larger than the highest multiplicity event in the training data and events with fewer particles are padded with zeros. The superscript indicates the time step in the diffusion process. The particle IDs are one-hot encoded and the diffusion model is trained to generate 26 distinct types of hadrons that make up more than 99$\%$ of particles produced at this energy. 

To capture the various correlations present in the training data, the diffusion model takes a latent vector $z$ as an additional input. A PointNet based encoder and a normalizing flow based decoder \cite{kobyzev2020normalizing}, trained end-to-end with the diffusion model, learn to map each UrQMD event into a latent vector $z$ that encodes various global features and correlations of the event. During generation, the normalizing flow decoder produces $z$, which becomes the input condition for the diffusion model. This conditioning facilitates the generation of physically realistic point clouds that preserve essential multi-particle correlations and global event properties that would otherwise be absent, since the diffusion model generates particles independently to maintain the permutation invariance of the point cloud representation.

The forward diffusion process is realized through a Markov chain that progressively adds noise to each point in the point clouds over several time steps until the data is transformed into complete noise. The conditional probability distribution $q(\mathbf{x}^{(1:T)}_i | \mathbf{x}^{(0)}_i)$ for a sequence of states $\mathbf{x}^{(1:T)}_i =\{\mathbf{x}^{(1)}_i, \mathbf{x}^{(2)}_i, ...,\mathbf{x}^{(T)}_i\}$ generated through the diffusion steps starting from the initial point $\mathbf{x}^{(0)}_i$ in the point cloud, is modeled as the product of probabilities of all consecutive diffusion steps:  
\begin{equation}
q(\mathbf{x}^{(1:T)}_i | \mathbf{x}^{(0)}_i) = \prod_{t=1}^T q(\mathbf{x}^{(t)}_i | \mathbf{x}^{(t-1)}_i),
\end{equation}
where $q(\mathbf{x}^{(t)}_i | \mathbf{x}^{(t-1)}_i)=\mathcal{N}\left(\mathbf{x}^{(t)}_i \big| \sqrt{1 - \beta_t} \mathbf{x}^{(t-1)}_i, \beta_t \mathbf{I}\right)$ is the diffusion kernel which introduces noise to the point in the previous timestep, and is modeled as a Gaussian with a mean of $\sqrt{1 - \beta_t} \mathbf{x}^{(t-1)}_i$ and variance of $\beta_t \mathbf{I} $. Here $\beta_t$ is a  hyperparameter controlling the amount of noise added at each timestep and $\mathbf{I}$ is the identity matrix with dimensions matching the particle vector $\mathbf{x}_i$.
 \begin{figure}[t]
   \includegraphics[width=0.49\textwidth]{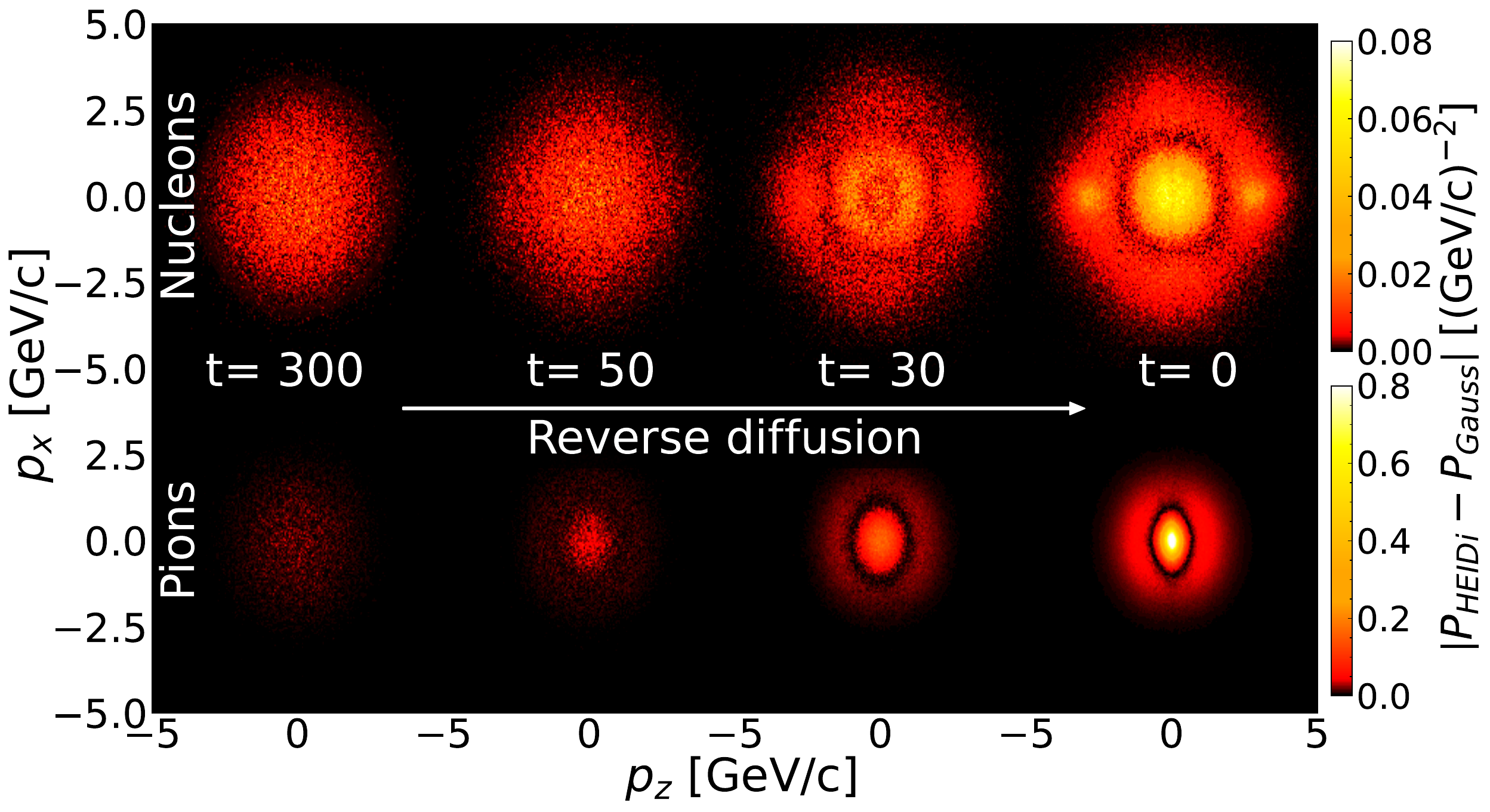}
   \caption{(Color online) Visualization of the generation process in \emph{HEIDi}. Starting from random initial values for the momentum and ID at timestep t=300, the reverse diffusion process progressively denoises the samples, ultimately producing realistic particles at t=0. The absolute deviation for the $p_x$-$p_z$ probability densities of \emph{HEIDi} from those of a random Gaussian probability density for various timesteps are shown. Note that \emph{HEIDi} generates particles, not distributions. The distributions presented are constructed from the particles in 2000 \emph{HEIDi} events.}
   \label{traj}
 \end{figure}
 \begin{figure}[t]
   \includegraphics[width=0.5\textwidth]{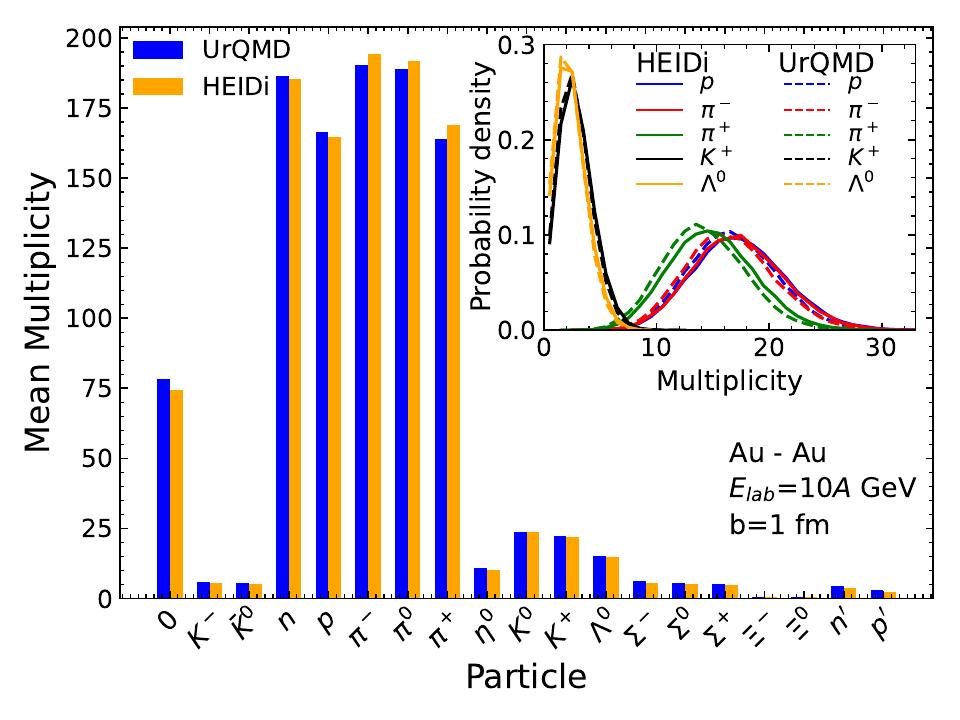}
   \caption{(Color online) Main figure: Mean event-multiplicity of various hadrons for Au-Au collisions at $10A$ GeV with b=1 fm. The results from the diffusion model are shown as orange bars, while the blue bars correspond to multiplicities from the UrQMD cascade model. The HEIDi model can also generate $\bar{p} ,\bar{n} , \bar{\Lambda},\bar{\Sigma}^-,\bar{\Sigma}^0,\bar{\Sigma}^+,\bar{\Xi}^0,\bar{\Xi}^+,\bar{\Omega}^+ \mathrm{ and }\ \Omega^-$ in an event. However, these hadrons have been excluded from the plot due to their very small multiplicities ($<$1/event). The particles marked \emph{0} are the fake particles used to maintain a constant multiplicity to the event while  \emph{n'} and \emph{p'} are spectator neutrons and protons respectively. Inset: Multiplicity distributions of selected hadrons at mid rapidity ($|y| < 0.1$). The results from the diffusion model are shown as solid lines, while the dashed lines correspond to multiplicities from the UrQMD cascade model.}
   \label{meanmulti}
 \end{figure}
 
The process of generating point clouds involves a reverse Markov chain, which takes as input the latent vector $z$ generated by the normalizing flow model, along with random samples from a simple noise distribution $\tilde{q}(\mathbf{x}^{(T)}_i) $ which approximates $q(\mathbf{x}^{(T)}_i)$. The Markov chain then sequentially denoise the random samples to generate states $\mathbf{x}^{(T-1)}_i, \mathbf{x}^{(T-2)}_i$,..., $\mathbf{x}^{(0)}_i$, ultimately recovering the original point cloud $\mathbf{x}^{(0)}_i$. The probability $\tilde{q}_\theta(\mathbf{x}^{(0:T)}_i | z) $ for a sequence of states $\mathbf{x}^{(0:T)}_i =\{\mathbf{x}^{(0)}_i, \mathbf{x}^{(1)}_i, ...,\mathbf{x}^{(T)}_i\}$, given a latent vector $z$, is then given by the product of the probabilities of all consecutive reverse diffusion steps and the initial noise distribution $\tilde{q}(\mathbf{x}^{(T)}_i )$:
\begin{equation}
\tilde{q}_\theta(\mathbf{x}^{(0:T)}_i | z) = \tilde{q}(\mathbf{x}^{(T)}_i )\prod_{t=1}^{T} \tilde{q}_\theta(\mathbf{x}^{(t-1)}_i | \mathbf{x}^{(t)}_i, z).
\end{equation}
The reverse diffusion kernel $\tilde{q}_\theta(\mathbf{x}^{(t-1)}_i | \mathbf{x}^{(t)}_i, z) $ is parameterized by a neural network with parameters $\theta$. This network is trained to learn the noise added in the previous timestep and to remove it at each step in the reverse diffusion process. The reverse diffusion kernel is thus also modeled as a Gaussian distribution given by
\begin{equation}
\tilde{q}_\theta(\mathbf{x}^{(t-1)}_i | \mathbf{x}^{(t)}_i, z) = \mathcal{N}(\mathbf{x}^{(t-1)}_i | \mu_\theta(\mathbf{x}^{(t)}_i  , t, z), \beta_t I),
\end{equation}
where $\mu_{\theta}$ is the de-noised mean as predicted by the network based on the current state $\mathbf{x}^{(t)}_i$, the time step $t$, and the latent vector $z$. This event generation process in \emph{HEIDi} is illustrated  in figure \ref{traj}. For more details on \emph{HEIDi}'s network structure, see \cite{OmanaKuttan:2025fqc}.

After training, 50000 events were generated using both UrQMD and \emph{HEIDi} to evaluate the performance of the generative model. These tests aim to quantify the model's ability in capturing the underlying physics of the HIC events and produce realistic point clouds that are consistent with the features of UrQMD generated events. 

\emph{HEIDi} generates a list of particle vectors with 26 different hadrons species, representing the complete output of a collision event. The mean multiplicites of various hadrons from events generated by \emph{HEIDi} are compared to those from UrQMD in figure \ref{meanmulti}. Evidently, the model accurately learns the differences between the multiplicities of various hadron types in the training data.

The multiplicity distributions of the selected hadrons at mid-rapidity are shown in figure \ref{meanmulti} (inset). Although \emph{HEIDi} generates events with 26 distinct hadron types, for clarity and conciseness, in figure \ref{meanmulti} (inset) and in subsequent plots, we show the results only for protons ($p$), charged pions ($\pi^-,\pi^+$) , Kaons ($K^+$) and Lambdas ($\Lambda^0$), which represent the most abundant hadrons in the system. It shows that the model, besides learning to reproduce mean multiplicities, can also effectively capture the probability distributions of multiplicities for various hadron species produced in an event. From less abundant particles, such as $\Xi$ or $\Sigma$, with multiplicities of $\approx$1-3 per event, to the most abundant particles, like protons and pions, \emph{HEIDi} accurately reproduces the UrQMD multiplicity distributions. 

\begin{figure}[t]
   \includegraphics[width=0.5\textwidth,keepaspectratio]{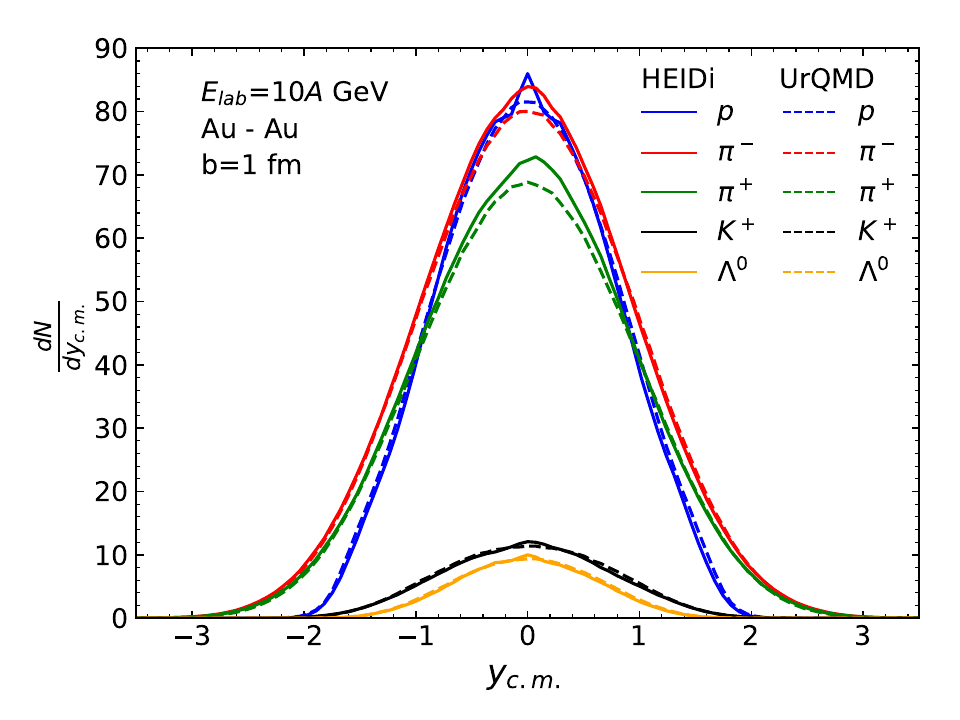}
   \caption{(Color online) Rapidity distributions of selected hadrons for Au-Au collisions at $10A$ GeV with b=1 fm. The solid lines show the diffusion model results while the dotted lines denote UrQMD results.}
   \label{rap}
 \end{figure} 
Figure \ref{rap} compares the rapidity distributions of the selected hadrons in the generated events with the corresponding distributions from UrQMD. The rapidity distributions of particles generated from \emph{HEIDi} are in good agreement with those simulated from the UrQMD for all hadron types. For high multiplicity particles like nucleons and pions, a small deviation appears around rapidities close to 0, where the generative model slightly overestimates the number of particles. This deviation is primarily due to a slight excess of the very low-momentum particles in the data generated by the diffusion model.
 
\begin{figure}[t]
   \includegraphics[width=0.5\textwidth]{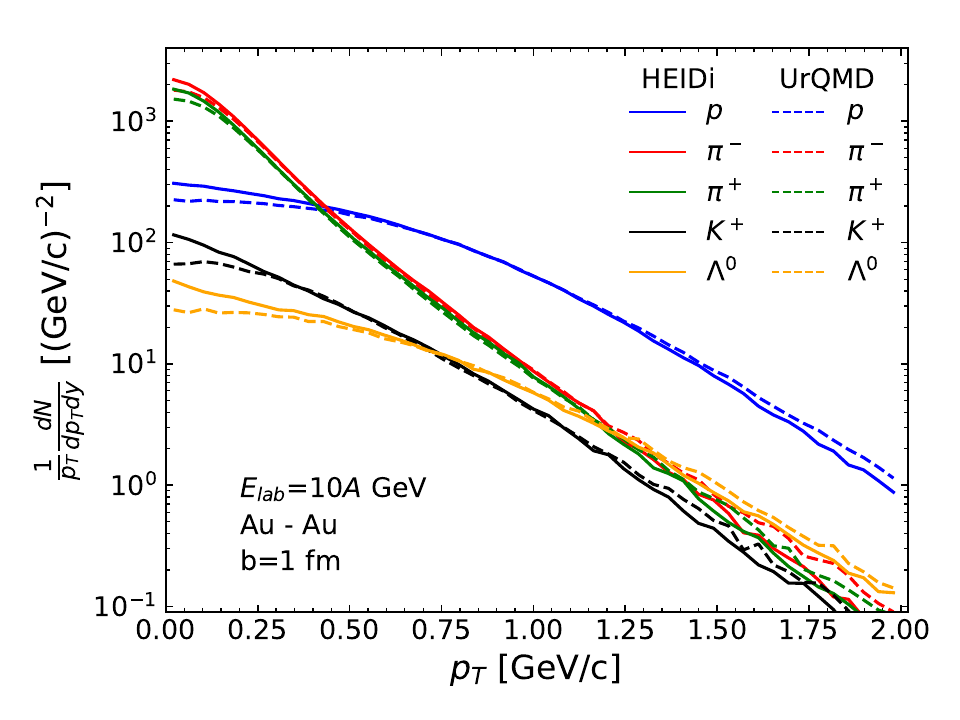}
   \caption{(Color online) Transverse momentum distributions of selected hadrons at mid rapidity ($|y| < 0.1$) for Au-Au collisions at $10A$ GeV with b=1 fm. The results of generative model are shown in solid lines while the dotted lines denote UrQMD results. The generative model successfully reproduces the UrQMD $p_T$ spectra across most momentum ranges, with the exception of the deviation at very low $p_T$ values.}
   \label{pt}
 \end{figure}

This effect is also noticeable in the transverse momentum ($p_T$) distributions at mid-rapidity ($|y| < 0.1$), as shown in Figure \ref{pt}. The particles generated by the generative model closely reproduces the $p_T$ distribution across most momentum ranges, but at lower $p_T$ values, $p_T <$ 400 MeV, $\emph{HEIDi}$ generates an excess of low-momentum particles over the UrQMD model. However, this large deviation at small $p_T$ values results only in a small difference to the total yield. This discrepancy suggests that the model does not fully capture the correlations in the very low-momentum region of phase space. While this issue could be mitigated using a more diverse training dataset, such as minimum-bias data, it is of less concern for experimental applications. Detectors typically have limited acceptance and efficiency for low-momentum particles \cite{STAR:2021yiu,HADES:2020wpc}. Hence, a low momentum cutoff is usually applied to the simulation data for analyses \cite{Reichert:2025rnw,Hillmann:2019wlt}. 

Overall, the present model does an impressive job in learning the probability distributions of various hadrons spanning four orders of magnitude, from abundant lower momentum particles to the exponentially less likely high momentum particles. To achieve this level of accuracy with a limited training dataset in point-cloud format is not trivial.
 
\begin{figure}[t]
   \includegraphics[width=0.5\textwidth,height=0.2\textheight]{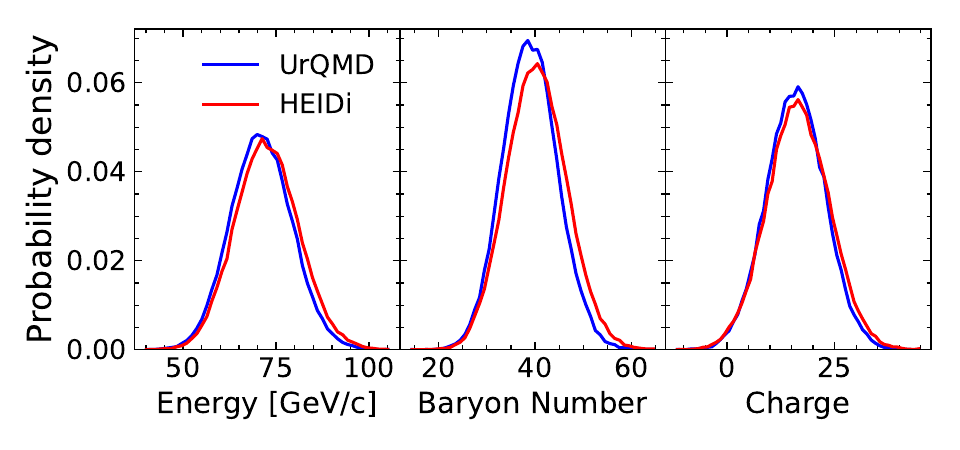}
   \caption{(Color online) Global features of generated events. The distributions of the total energy, baryon number and charge of all mid rapidity particles ($|y| < 0.1$) are shown in panels 1, 2 and 3, respectively. Results from the generative model are shown in red, while the corresponding UrQMD distributions are shown in blue.}
   \label{ene}
 \end{figure}

Besides single particle specific quantities like momenta and multiplicities of various hadrons, it is also important to check how well global, event-by-event quantities like total energy ($E$), total baryon number ($B$) and total charge ($Q$), as well as dynamic correlations like collective flow are learned by the generative model. These results are presented in figure \ref{ene}, where the distributions of the $E$, $B$ and $Q$ values of all generated hadrons at mid rapidity per event are compared to those obtained from UrQMD calculations. Notably, such global features are also well captured by the diffusion model. The AI generated events and UrQMD generated events have very similar distributions for $E$, $B$ and $Q$. Furthermore, the generative model also accurately reproduces the vanishing elliptic flow for protons and pions, consistent with the UrQMD results at mid-rapidity for central, $b$=1 fm collisions at $10A$ GeV. The model provides sufficiently accurate outputs for most event averaged observables. However, as evident in figure \ref{ene}, observables sensitive to strict conservation laws such as event-by-event fluctuation measurements may require further refinements to the model or additional conditioning strategies. For a detailed analysis of \emph{HEIDi}'s performance, see \cite{OmanaKuttan:2025fqc}.

While accurately capturing the physics and correlations observed in UrQMD is essential, a primary motivation for implementing deep learning models is its significant speedup, particularly when the model is deployed on GPUs. On a Nvidia A100 GPU with 40 GB of memory, the model generates an event in about 30 milliseconds while the UrQMD model takes about 3 seconds to generate a single event. This represents a speedup of two orders of magnitude. It is important to note that here, UrQMD in cascade mode, the fastest version, was used for demonstration. The inclusion of potential interactions in UrQMD will increase the computational time to one minute, and the hybrid version of UrQMD, which includes an intermediate hydrodynamic stage, can take up to one hour per event at SIS-100 energies. When trained on such data, the deep learning model will deliver a speedup of at least five orders of magnitude.

The capability of \emph{HEIDi} to not only distinguish a wide range of hadron species, but also to learn well their multiplicity and momentum distributions, while correctly capturing various global features of the event, marks a significant advancement in generative modeling of heavy-ion collisions. Nevertheless, practical use in downstream tasks such as Bayesian inference, detector-level simulations, and data-driven physics analyses requires the model to generalize across a broader set of conditions, including varying centralities, beam energies, collision systems, EoSs and various detector effects. The network architecture \cite{OmanaKuttan:2025fqc}, the point cloud  data format and the limited training dataset were chosen to support this adaptability. 

Extending \emph{HEIDi} to incorporate and generalize across various conditions or physical parameters is straightforward. Any additional conditioning variable can be introduced to the model by appending it to the latent vector $z$, which is generated by the normalizing flow-based decoder. A proof-of-principle demonstration of this adaptability, without requiring any changes to the network architecture, is presented in the Appendix \ref{appendix}. This highlights that extending the model to other physical parameters is straightforward, requiring only the inclusion of those parameters in the conditioning vector and appropriate training data.

Such an ultra-fast point-cloud diffusion model for heavy-ion collisions could eliminate the need for training specific surrogate models to predict individual observables, as the model is capable of generating events at high speed and directly calculating any observable. This will make it an indispensable tool for comprehensive Bayesian analysis. Furthermore, a deep generative model can serve as a differentiable version of the physics model which can be directly used for parameter estimation tasks or to perform comprehensive global fits directly on raw, particle level experimental data. This opens the door to novel gradient-based inference methods in heavy-ion physics for the first time. This approach eliminates the need to train deep learning models for specific tasks or inputs and improves interpretability, as the result can then be directly verified on the original model.  

The point cloud structure of electronically collected data in sensors also offers exciting potential for point cloud diffusion models in biomedical imaging and diagnostics, natural hazard early warning systems, autonomous navigation, robotics etc. In high-energy physics, the presented model is flexible to be adapted for any event-by-event model calculations, experimental simulations, or a combination of both. It is more precise and avoids the loss of information typically associated with histograms or image-based representations. Due to the compact representation of point clouds, these models can also be more resource-efficient as compared to generating images or voxel grids. Both the theoretical and experimental communities can rely on point cloud diffusion models to quickly generate large numbers of events for initial-level analysis, the results of which can later be verified using actual physics models. 

Even though the presented results are highly promising, extensive tests on the reliability and consistency of the model are still necessary before HEIDi can be applied to physics analyses. In particular, the sources of the deviations observed at small transverse momentum and rapidities need to be better understood, and further developments are required to mitigate these discrepancies. Moreover, it would be valuable to compare and quantitatively benchmark the performance of HEIDi against other point-cloud–based generative models developed for particle physics tasks. Such comparisons could help identify the limitations of PointNet-based architectures in capturing complex higher-order correlations in the data. However, these efforts lie beyond the scope of the present work and represent important directions for future research. While not yet a complete foundation model, the presented \emph{HEIDi} framework represents a key step toward building one for heavy-ion physics through a fast, general-purpose, and extensible full event-by-event generation method adaptable for a wide range of downstream tasks.

\begin{acknowledgements}
 This work is supported by the the BMBF under the KISS project (M.O.K, K.Z), the SAMSON AG (J.S, K.Z), the CUHK-Shenzhen university development fund under grant No. UDF01003041 and UDF03003041 (K.Z), and Shenzhen Peacock fund under No. 2023TC0179 (K.Z) and the Walter Greiner Gesellschaft zur F\"orderung der physikalischen Grundlagenforschung e.V. through the Judah M. Eisenberg Laureatus Chair at the Goethe Universit\"at Frankfurt am Main (H.S).
\end{acknowledgements}

\appendix 
\section{Adaptability of the model}
\label{appendix}
To demonstrate the easy adaptability of the model  across various conditions or physical parameters, \emph{HEIDi} was trained on 30000 UrQMD events with different collision centralities. The dataset consisted of 10000 events each at impact parameters of 1 fm, 3 fm, and 5 fm. We used the same network architecture as in the main results. The only modification was the inclusion of the impact parameter, concatenated with the latent vector $z$, as an additional conditioning input to the diffusion model.

\begin{figure}[!htpb]
 \centering
\includegraphics[width=0.54\textwidth,keepaspectratio]{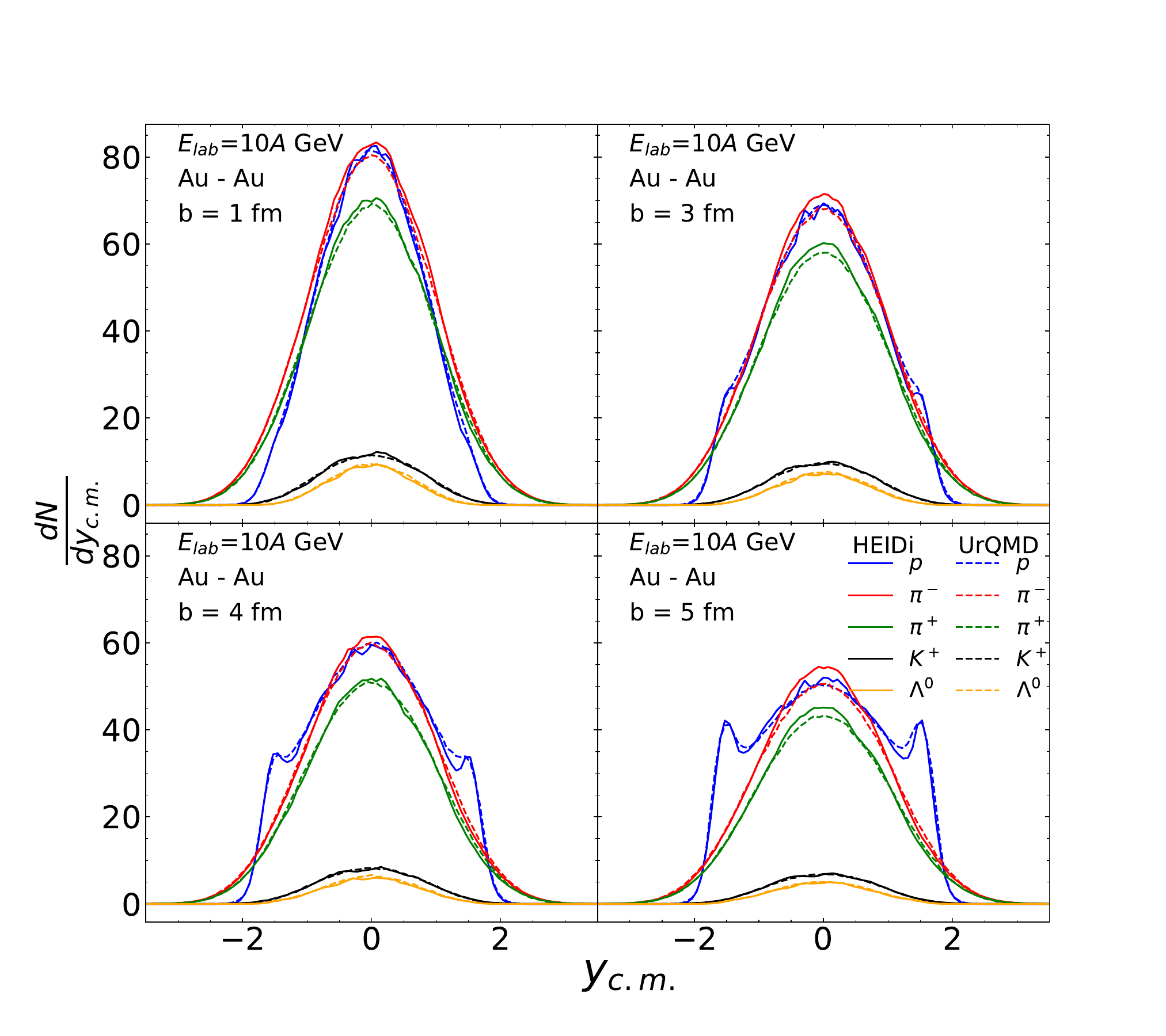}
   \caption{Rapidity distributions of selected hadrons in Au–Au collisions at $10A$ GeV for impact parameters b = 1 fm (top left), 3 fm (top right), 4 fm (bottom left) and 5 fm (bottom right). Solid lines represent predictions from \emph{HEIDi}, while dotted lines show the UrQMD results. The model was trained only on b = 1, 3, and 5 fm, yet it accurately reproduces the distributions not only for the trained impact parameters but also for b= 4 fm (bottom left panel), which was not included in the training data.}
   \label{rap_b}
    \end{figure}

The model trained in this way can generate realistic collisions conditioned on different impact parameters. Figure \ref{rap_b} shows the rapidity distributions of selected hadrons at various impact parameters. It can be seen that the resulting model can now generate realistic collisions for varying centralities.
The lower left panel corresponds to the results for b = 4 fm, a value not present in the training set. The model also generalizes well to this unseen impact parameter. 
This proof of principle highlights the flexibility and extensibility of the \emph{HEIDi} framework. Further conditioning or adaptations of the model on physical parameters such as beam energy, equation of state, collision system etc., as well as the inclusion of various detector effects, requires only appropriate training data and does not require any modification to the underlying architecture.

\FloatBarrier

\bibliography{Bl_EoS}
\bibliographystyle{apsrev4-1}

\end{document}